\documentclass[conference]{IEEEtran}

\IEEEoverridecommandlockouts

\ifCLASSOPTIONcompsoc
  \usepackage[nocompress]{cite}
\else
  \usepackage{cite}
\fi

\usepackage{amsmath}

\usepackage{fixltx2e}
\usepackage{stfloats}

\usepackage{url}

\usepackage{listings}
\usepackage{graphicx}
\usepackage{paralist}
\usepackage{amssymb}
\usepackage{xcolor,colortbl}

\usepackage{pifont}
\usepackage[ruled,linesnumbered,vlined]{algorithm2e}

\renewcommand{\paragraph}[1]{\vspace{1ex}\noindent
  \textbf{\textit{#1}.}\hspace{1ex}}

\newcommand{\sysname}{$\mu$Dep}

\begin{document}

\title{\sysname{}: Mutation-based Dependency Generation for Precise Taint Analysis on Android Native Code}

\author{Cong~Sun, Yuwan~Ma, Dongrui~Zeng, Gang~Tan, Siqi~Ma, and~Yafei~Wu 

\thanks{Cong~Sun, Yuwan~Ma, and Yafei~Wu are with the School of Cyber Engineering, Xidian University, China.}
\thanks{Dongrui~Zeng was with the Pennsylvania State University and is now with Palo Alto Networks.}
\thanks{Gang~Tan is with the Pennsylvania State University.}
\thanks{Siqi~Ma is with the University of New South Wales, Australia.}
\thanks{The first two authors contribute equally to this work.}
\thanks{Corresponding author: Cong~Sun. E-mail: suncong@xidian.edu.cn.}
\thanks{Cong Sun, Yuwan~Ma, and Yafei~Wu were supported by the National Natural Science Foundation of China (No. 61872279) and the Key Research and Development Program of Shaanxi (No. 2020GY-004).}
}

\maketitle

\begin{abstract}
The existence of native code in Android apps plays an important role in triggering inconspicuous propagation of secrets and circumventing malware detection. However, the state-of-the-art information-flow analysis tools for Android apps all have limited capabilities of analyzing native code. Due to the complexity of binary-level static analysis, most static analyzers choose to build conservative models for a selected portion of native code. Though the recent inter-language analysis improves the capability of tracking information flow in native code, it is still far from attaining similar effectiveness of the state-of-the-art information-flow analyzers that focus on non-native Java methods. To overcome the above constraints, we propose a new analysis framework, \sysname{}, to detect sensitive information flows of the Android apps containing native code. In this framework, we combine a control-flow based static binary analysis with a mutation-based dynamic analysis to model the tainting behaviors of native code in the apps. Based on the result of the analyses, \sysname{} conducts a stub generation for the related native functions to facilitate the state-of-the-art analyzer DroidSafe with fine-grained tainting behavior summaries of native code. The experimental results show that our framework is competitive on the accuracy, and effective in analyzing the information flows in real-world apps and malware compared with the state-of-the-art inter-language static analysis.
\end{abstract}


\section{Introduction}\label{sec:introduction}

Native code is widely embedded in Android apps to benefit code reuse and processor-intensive tasks, e.g., video processing and game graphics engine. Recent statistics showed that around 36\%-40\% of regular android apps make use of native code \cite{DBLP:conf/ndss/AfonsoGBFKVDP16, DBLP:conf/ccs/WeiLOCZ18}. However, native code is also vulnerable to the exploitations by the recent Android malware because of its ``black-box'' feature to escape traditional static analyses of malicious behaviors. For instance, native code can be used to hide sensitive data or code stub \cite{DBLP:conf/ndss/RasthoferAMB16, DBLP:conf/wistp/RasthoferAHB15}, conduct full-native code obfuscation \cite{DBLP:conf/uss/WongL18}, tamper with Dalvik's data memory \cite{DBLP:conf/ndss/SeoKCSK16}, or circumvent the state-of-the-art static data-flow analysis \cite{DBLP:conf/ccs/WeiLOCZ18}. In these forms of malicious exploits, leaking sensitive information is one of the major threats conducted by native code \cite{DBLP:conf/dsn/QianLSC14, DBLP:conf/uss/XueZCLG17, DBLP:conf/ccs/WeiLOCZ18}. Specifying the data-flow behaviors of native code becomes the key step to identify such kind of threats.

To model the tainting behaviors of native code, the dynamic approaches \cite{DBLP:conf/osdi/EnckGCCJMS10, DBLP:conf/uss/YanY12} applied conservative strategies to propagate taints through either JNI method calls or instructions at the machine-code level. On the other hand, state-of-the-art static information flow analyses on Android apps \cite{DBLP:conf/pldi/ArztRFBBKTOM14, DBLP:conf/ccs/WeiROR14, DBLP:conf/ndss/GordonKPGNR15, DBLP:conf/eurosp/CalzavaraGM16} only inspect data flows at the bytecode level but fail to analyze the native code in the apps. The static analysis usually builds summaries to specify the data-flow behaviors for only a limited number of native methods. For example, the native call handler of FlowDroid \cite{DBLP:conf/pldi/ArztRFBBKTOM14} hard-codes the models of taint propagation for several system-defined native methods. If the input is tainted before such a native call, the taint is propagated to all the arguments and its return value. DroidSafe \cite{DBLP:conf/ndss/GordonKPGNR15} uses a manually crafted comprehensive execution model, called \emph{accurate analysis stubs}, to specify runtime behaviors including data flow and object instantiation of native code in the Android framework. This model over-approximates the taint propagation of standard Android native libraries, but fails to characterize the tainting behaviors of user-defined or third-party native code. The potential propagations passing through these native codes are simply cut off by the model. StubDroid \cite{DBLP:conf/icse/ArztB16} automatically summarizes the tainting behaviors of Android framework libraries in bytecode to improve the efficiency of data-flow analysis. However, the summaries of native code are still built manually.

In general, the vulnerable behaviors of native code that threaten the information flow security can be classified into two types. First, on the native side, the binaries of JNI functions may invoke sources or sinks to track or release sensitive data, which we call \emph{type-I} vulnerabilities. The sources and sinks may be either Java methods or native library functions. Second, there are many cases that the leakages arising on the bytecode are triggered by some inconspicuous propagations made by native code, which we define as \emph{type-II} vulnerabilities. White-box analysis of the native code may help identify such vulnerabilities. Lantz et al. \cite{DBLP:conf/iwcmc/LantzJ15} proposed identifying sources and sinks called in native libraries by traversing the program dependency graph (PDG) of native function constructed with IDA. They focused on type-I vulnerabilities. JN-SAF \cite{DBLP:conf/ccs/WeiLOCZ18} is the first approach to capturing the inter-language data flows. A summary-based bottom-up approach based on \cite{DBLP:conf/pldi/DilligDAS11} is used to perform flow- and context-sensitive inter-language data-flow analysis, whose summaries are generated to unify the heap manipulations of both Java bytecode and native code. Although addressing both type-I and type-II vulnerabilities, the analysis of native code is still not precise to capture the taint propagations conducted by native code.

To facilitate the static information flow analysis with precise specifications of inconspicuous sensitive data flows in native code, we propose a hybrid approach to automatically build more precise tainting behavior models for the native code of Android apps. Firstly, we develop a lightweight static binary analysis to deal with type-I vulnerabilities and abstract the tainting effects of natively called sources/sinks to the bytecode. Then, we leverage the principle of differential fuzzing \cite{DBLP:conf/icse/NilizadehNP19} to trigger argument mutations over self-composed invocations of the native method and identify the dependencies between arguments and return values of the native method. Based on the dependency relations, we propose automatically generating stubs to improve the precision of the Android Device Implementation (ADI) model of DroidSafe \cite{DBLP:conf/ndss/GordonKPGNR15}. The newly generated code stubs precisely model the data-flow effects of user-defined or third-party native code, and provide mechanisms for the analyzer to identify type-II vulnerabilities.
We highlight our contributions as follows:

\begin{itemize}
\item We propose \sysname{}, an information flow analysis framework tightly integrated with DroidSafe \cite{DBLP:conf/ndss/GordonKPGNR15} to identify sensitive data flows in the native code of apps. It combines a control-flow based static binary analysis and a mutation-based dynamic analysis to specify the tainting effects of the native code in the bytecode of app.

\item The stub generation procedure uses mutation-based dependencies to build precise data-flow models for the native code. The derived semantics of the native code are merged with the ADI model of DroidSafe to implement a more precise information flow analysis.

\item The experimental results show that our information flow analysis framework is competitive on accuracy compared with the state-of-the-art framework JN-SAF \cite{DBLP:conf/ccs/WeiLOCZ18}, and demonstrate the effectiveness in detecting sensitive information flows in real-world apps.
\end{itemize}

This paper is organized as follows. We first present a motivating example in Section~\ref{sec:motivating_example}. Section~\ref{sec:design} describes the design of our approach. Section~\ref{sec:implementation} discusses the implementation issues. The evaluation and security findings are presented in Section~\ref{sec:evaluation}. Section~\ref{sec:discuss} makes several discussions. Section~\ref{sec:related-work} presents the related works and we conclude this paper in Section~\ref{sec:conclusion}.

\section{Motivating Example}\label{sec:motivating_example}

\begin{figure}[t]
  \centering
  \includegraphics[width=\linewidth]{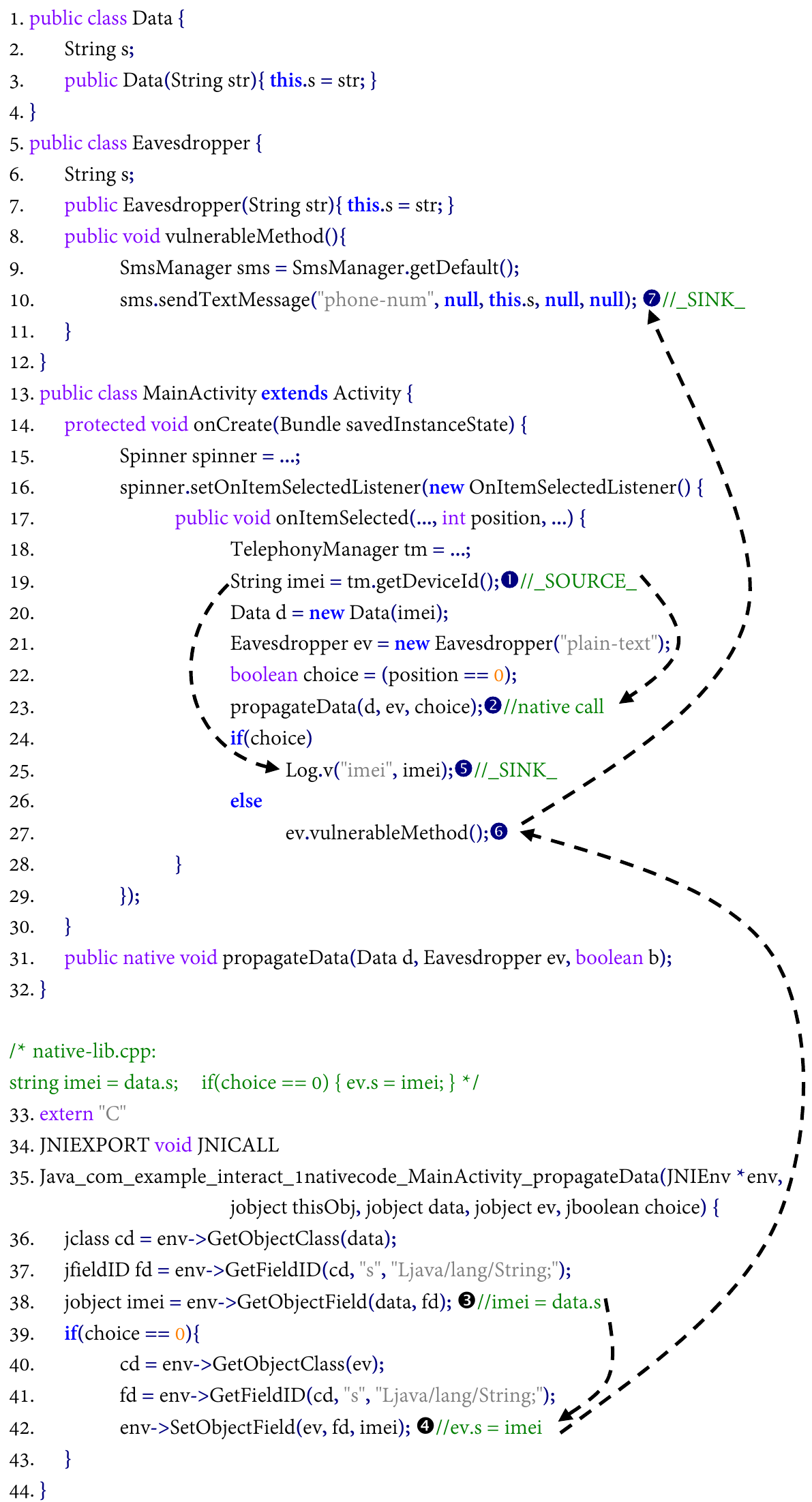}
  \caption{Motivating Example}\label{fig:example}
\end{figure}

In this section, we motivate our solution by presenting an example (in Figure~\ref{fig:example}) to show the deficiency of two existing state-of-the-art information flow analyzers DroidSafe~\cite{DBLP:conf/ndss/GordonKPGNR15} and JN-SAF \cite{DBLP:conf/ccs/WeiLOCZ18}.

The example application uses \verb|TelephonyManager| to get the local IMEI when the user operates on a Spinner component. The IMEI is delivered to the native method \texttt{propagateData} through an object of \texttt{Data}, i.e. \ding{182}$\rightarrow$\ding{183}, where an error-prone instance of \verb|Eavesdropper| and a GUI option \verb|choice| are also passed. From the definition of \verb|Eavesdropper| we cannot identify malicious behavior except its field \verb|s| is sent to some phone in text message. However, in the native function of \verb|propagateData|, the IMEI held by \verb|Data| object is assigned to the field \verb|s| of \verb|Eavesdropper| under certain input, i.e. \verb|choice| is \verb|false| (0 in native code), i.e. \ding{184}$\rightarrow$\ding{185}. Then, after returning to the Java side, the IMEI is sent to another phone in method \verb|vulnerableMethod| through text message, i.e. \ding{185}$\rightarrow$\ding{187}$\rightarrow$\ding{188}. On the other hand, if the input \verb|choice| is \verb|true|, the IMEI will be printed to log.

To analyze the information flows of this app, we assume \verb|TelephonyManager.getDeviceId()| is the sensitive source. \verb|SmsManager.sendTextMessage()| and \verb|Log.v()| are public sinks. DroidSafe can only build a mock implementation for the native method \verb|propagateData|, which will cut off the dependency between \verb|Data| and \verb|Eavesdropper| on the native side. Consequently, it can only detect a sensitive flow \ding{182}$\rightarrow$\ding{186}, but fail to detect another flow \ding{182}$\rightarrow$\ding{183}$\rightarrow$\ding{184}$\rightarrow$\ding{185}$\rightarrow$\ding{187}$\rightarrow$\ding{188} causing type-II vulnerability. Meanwhile, possible implementation limitations of JN-SAF may have caused heap-manipulation summary missing, while applying the summary of \verb|vulnerableMethod()| to the native side of \verb|propagateData| in the example, resulting in only detecting \ding{182}$\rightarrow$\ding{186}. In contrast, we provide an automated approach to figure out native-side dependencies and generate informative and precise summaries for native methods.
Then the static information flow analysis can better bridge the contexts before and after the native calls and detect both sensitive flows in this example.

\section{Design of \sysname{}}\label{sec:design}

In this section, we describe the design of our information flow analysis framework of Android apps with the ability to detect type-I and type-II vulnerabilities caused by native code.

\subsection{Overview}

\begin{figure}[t]
  \centering
  \includegraphics[width=\linewidth]{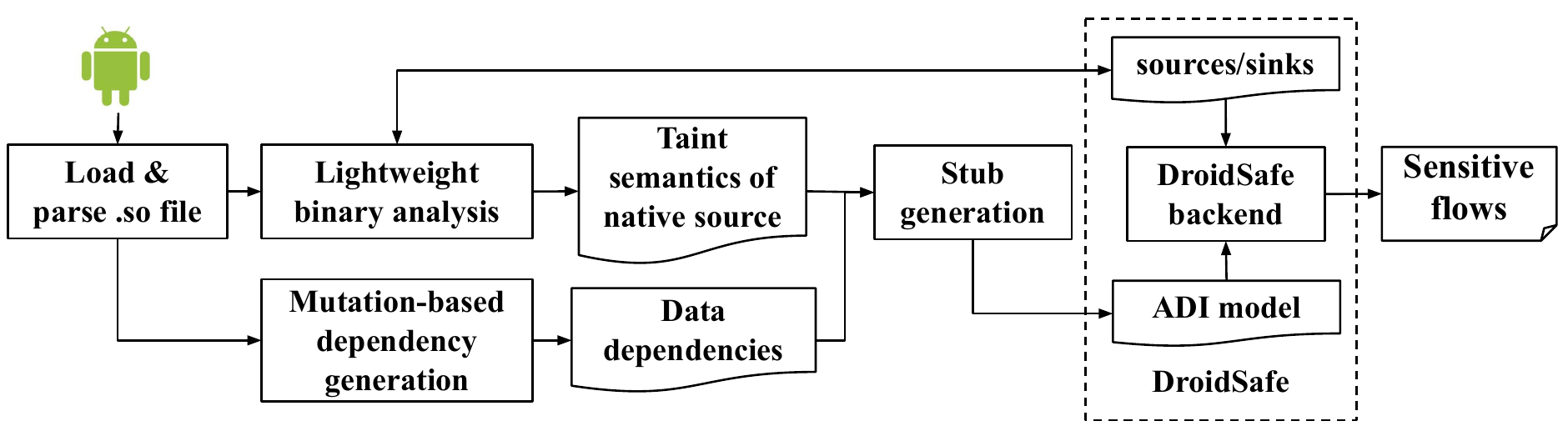}
  \caption{Framework of \sysname{}}\label{fig:framework}
\end{figure}

We show the framework of \sysname{} in Figure~\ref{fig:framework}. Firstly, \sysname{} loads the shared object files of each app and conducts a lightweight static binary analysis based on IDA \cite{idapro}. The binary analysis parses the sequences of JNI function calls in the body of each native function. Based on the pattern of JNI function call sequences, we find the Java APIs called in native code. Then \sysname{} checks if the found method call is a source or a sink on the control-flow graph of the native methods. If so, we conservatively identify such a control-flow relation as a type-I vulnerability. We then update the sources or sinks of DroidSafe, over-approximate the semantics of natively called source to the Java side, and deliver such semantics to the stub generation procedure.

Then, we develop a mutation-based dependency generation to build all the data dependencies between the arguments and the return value of each native method used in the app. To reach a reasonable granularity, the fields of arguments and returned object on different depths are considered.

Thirdly, \sysname{} generates the code stubs in Jimple automatically for the user-defined or third-party native methods based on the data dependencies as well as the taint semantics of the natively called source. The stubs extend the ADI model of DroidSafe and provide the model for the inconspicuous taint propagation behaviors of native functions, which is critical for the DroidSafe engine to detect the type-II leakages.

\subsection{Lightweight Static Binary Analysis}\label{subsec:ida}

To deal with the type-I vulnerability that the sources or sinks are invoked in native code, we developed a lightweight control-flow based binary analysis with IDA. The principle of our analysis is to use the native method itself as a proxy to represent the source or sink called by the corresponding native function. Through updating the source or sink list, the information flow analysis can then detect type-I vulnerabilities. The steps of binary analysis are as follow.

\begin{compactenum}[1.]
\item We retrieve the code section of shared object files, find the calls to the specific JNI functions, e.g., \verb|CallXXXMethod|, and decide if the pattern of related instructions is calling some Java source or sink method. We also find the calls to native sources and sinks which are some native library functions, e.g., \verb|__android_log_print|.

\item From the call sites of these retrieved sources and sinks, we traverse backward along the binary-level control-flow graph to find the native functions that use at least one of these sources and sinks.

\item For each of the identified native functions, if there is a corresponding native method declared at the Java side of the app, we find a coarse-grained correlation representing the usage of source/sink by the native method. Then we add the native method into the source or sink list of DroidSafe.
\end{compactenum}
To map the native function with its native method on the Java side, we have to resolve the dynamic function registrations conducted by calling \verb|RegisterNatives| in \verb|JNI_OnLoad|. A primary argument of \verb|RegisterNatives| is a list of \verb|JNINativeMethod| structures usually stored in the data sections. For the initial address $addr$ of each native function that uses source or sink, we search for all the occurrences of $addr$ in the data sections, take this $addr$ as the third field of \verb|JNINativeMethod| structure and resolve the first and second field of address that point to the string of native method name and native method signature respectively. Based on the resolved method name and method signature, we update the source or sink list of DroidSafe.

We take into account the return type of the natively called source to build the taint semantics for the proxy source method added into the source list, which will then be merged with the data dependencies derived in Section~\ref{subsec:dep_gen} to construct the code stub of the native method. The strategy is, if the return type of natively called source is $T$, then for the return value and output arguments of the proxy native method, we search for and taint all their fields with type $T'$ compatible with $T$.  To overcome the constraint that DroidSafe can only taint the returned object and the output arguments of proxy source method instead of their fields, we build a corresponding Java wrapper method to first call the proxy source method and then taint all the related fields of its return object or output arguments. We substitute the call to the proxy source method with a call to this wrapper method.

\subsection{Mutation-based Dependency Generation}\label{subsec:dep_gen}

We generalize the idea of \emph{differential fuzzing} \cite{DBLP:conf/icse/NilizadehNP19} and self-composition \cite{DBLP:conf/csfw/BartheDR04} to design a dynamic approach for generating the dependency relations between arguments and return values of native methods. The idea is to treat a native method as a black-box system and compare two executions of the system with inputs mutated to observe differences in outputs. As a result, we can infer the outputs that are changed should have dependency on the inputs that are mutated. For a native method, we consider the arguments with \texttt{String} or primitive types as input, the arguments referencing non-primitive-type objects as both input and output, and the return value as output. In detail, we iteratively mutate each input argument and observe the outputs. Since only one input is mutated in each iteration, if the coupled outputs of the two executions are different, the difference is guaranteed to be caused by the input change, which demonstrates the dependency relation between the mutated input and the altered output. In practice, by mutating each input argument multiple times, most of the paths are covered, and the dependencies between inputs and outputs are usually found.

To simplify the discussion of our algorithm, we define an \emph{operation unit} as two executions of one native method taking different inputs. For implementation simplicity to avoid possible parallel executions, in each operation unit, we take the principle of self-composition to compose a call to a native method sequentially with another call to the same native method. For the two calls to the same native method in one operation unit, we need to make sure their non-primitive-type arguments and return values are stored in distinguished memory locations so that the second execution would not override the first execution's output. Before we demonstrate the dependency generation algorithm, we first define several predicates to clarify the atomic operations used in this algorithm.

\begin{compactenum}[1.]
\item $\textit{clone}_{T}(ref, ref')$: Deep clone the object referenced by $ref$, return the reference $ref'$ to the newly generated object. Both objects have type $T$.

\item $\textit{cmp}_{T}(ref, ref')$: Deep compare the two objects referenced by $ref$ and $ref'$, both having type $T$. If $T=T'[]$, the length of array should also be compared in addition to the element comparisons.

\item $\textit{mutate}_{T}(ref, ref')$: Similar to the procedure of $\textit{clone}_{T}(ref, ref')$, except that whenever constructing an object for $ref'$ or some field of the new object, we feed randomized primitive-type value or object with immutable non-primitive types, or recursively apply the mutation operation on mutable non-primitive-type fields or some array element. The procedure is defined in Algorithm~\ref{algo:mutation}.
\end{compactenum}

\IncMargin{1em}
\begin{algorithm}[t]\footnotesize
\caption{$\textit{mutate}_T(ref,ref')$}\label{algo:mutation}

\uIf{\textit{isPrimitive}$(T) \vee $\textit{immutableNonPrimitive}$(T)$}{
    $ref'\leftarrow$ random value/object\;
}
\uElseIf{$T=T'[]$}{
    Alloc space for $ref'$, randomly select $i\in [0,ref.len)$\;
    \textit{mutate}$_{T'}(ref[i],ref'[i])$\;
    \For{$k \in [0,ref.len) \wedge k\neq i$}{
        \textit{clone}$_{T'}(ref[k],ref'[k])$\;
    }
}
\Else(\tcc*[h]{$T$ is mutable non-primitive type}){
    Alloc object space for $ref'$\;
    Randomly select field $(fd_i:T_i) \in T$\;
    \textit{mutate}$_{T_i}(ref.fd_i, ref'.fd_i)$\;
    \For{$(fd_k:T_k) \in T \wedge k\neq i$}{
        \textit{clone}$_{T_k}(ref.fd_k,ref'.fd_k)$\;
    }
}
\end{algorithm}\DecMargin{1em}

\begin{figure}[!t]
\centering
\includegraphics[width=\linewidth]{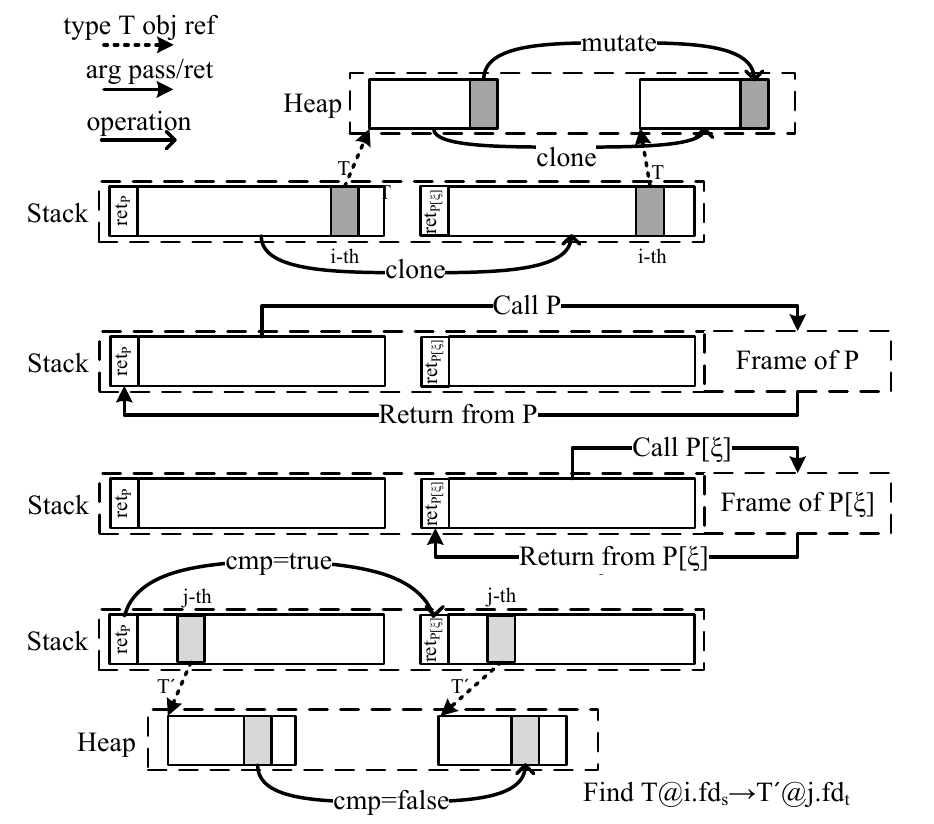}
\caption{Principle of Dependency Generation}
\label{fig_principle}
\end{figure}

Fig.~\ref{fig_principle} illustrates the variation on the memory (stack and heap) when executing one operation unit of our dependency generation approach. $P$ and $P[\xi]$ are both the execution of the same native method. $\xi$ is a renaming function that renames the arguments of $P$ syntactically to the new arguments used by $P[\xi]$. We compose $P$ and $P[\xi]$ sequentially as $P;P[\xi]$.  Then there are four states of memory in Fig.~\ref{fig_principle}: 1) \emph{before calling $P;P[\xi]$}, 2) \emph{calling $P$}, 3) \emph{calling $P[\xi]$}, and 4) \emph{after calling $P;P[\xi]$}.

In each operation unit, $P$ firstly takes the prepared arguments, and then $P[\xi]$ takes the cloned arguments as well as only one mutated argument. Each non-primitive-type argument of $P[\xi]$ holds a reference to the object generated by \textit{clone} or \textit{mutate}. For example, for the argument \verb|ev| of \texttt{propagateData()} in Fig.~\ref{fig:example}, there should be a renamed argument $\xi(\texttt{ev})$ which references the object generated by $clone_{\texttt{Eavesdropper}}(\texttt{ev}, \xi(\texttt{ev}))$ or $mutate_{\texttt{Eavesdropper}}(\texttt{ev}, \xi(\texttt{ev}))$. The instance of \texttt{MainActivity} delivered through the implicit argument \textit{this} should also be deeply cloned. We also store its reference onto the stack frame. After the sequential execution of the two calls, we apply $\textit{cmp}_T$ to compare the effect of $P$ and $P[\xi]$ on each pair of output arguments and return values. When the objects referenced by a pair of arguments, e.g., the $j$-th argument of $P$ and $P[\xi]$ in Fig.~\ref{fig_principle}, differentiate according to the result of $\textit{cmp}_T$, the diversity should either be caused by the mutation on the $i$-th argument, or by some indeterminate input from the native side. Then we build the dependency between the $i$-th argument and the $j$-th argument conservatively. We have also recorded the object fields that lead to the mutation of input and the diversity of output, e.g., in Fig.~\ref{fig_principle}, $fd_s$ is a field of the $i$-th argument with type $T$, and $fd_t$ is a field of the $j$-th argument with type $T'$. We finally generate the dependency bridging these fields, e.g., $T@i.fd_s\rightarrow T'@j.fd_t$ in Fig.~\ref{fig_principle}. The depth of object fields is configurable to support different granularity of the dependency relation.

The algorithm of the mutation-based dependency generation is given in Algorithm~\ref{algo:dependency_gen}. This algorithm assumes the instance native method $mtd$ has return type $\tau$. For each input argument under mutation, we repeatedly run fresh operation units for \textit{BOUND} times to reduce the false negatives of this dynamic approach. For the example in Fig.~\ref{fig:example}, we use this algorithm and derive the following dependency relation $\{$ $\langle\verb|Eavesdropper|@1.\verb|s|, \verb|Data|@0.\verb|s|\rangle,$ $\langle \verb|Eavesdropper|@1.\verb|s|, \verb|Boolean|@2\rangle\}$. Developing the dynamic analysis in this section as an app is critical to make the native code in shared object files runnable in the analysis, as stated in Section~\ref{sec:implementation}.

\IncMargin{1em}
\begin{algorithm}[t]\footnotesize
\caption{Mutation-based Dependency Generation}\label{algo:dependency_gen}

\DontPrintSemicolon
\SetKwData{Left}{left}\SetKwData{This}{this}\SetKwData{Up}{up}
\SetKwFunction{Union}{Union}\SetKwFunction{FindCompress}{FindCompress}
\SetKwInOut{Input}{input}\SetKwInOut{Output}{output}
\Input{native method $\tau\ T_0.mtd(T_1\ arg_1,\ldots, T_n\ arg_n)$}
\Output{dependency relation $\mathcal{D}$}
$\mathcal{D} \leftarrow \emptyset$\;
\For{$i=0$ \KwTo $n$}{
    \For{$time=0$ \KwTo \textit{BOUND}}{
        \For{$k=0$ \KwTo $n$}{
            Construct value/object randomly for $arg_k$\;
            \uIf{$k=i$}{
                $mutate_{T_k}(arg_k,\xi(arg_k))$\;
            }
            \uElseIf{$\textit{isPrimitive}(T_k)$}{
                $\xi(arg_k)\leftarrow arg_k$\;
            }
            \Else{
                $clone_{T_k}(arg_k,\xi(arg_k))$\;
            }
        }
        $ret\leftarrow arg_0.mtd(arg_1, \ldots, arg_n)$\;
        $ret'\leftarrow \xi(arg_0).mtd(\xi(arg_1), \ldots, \xi(arg_n))$\;
        \For{$(k=0$ \KwTo $n) \wedge (k\neq i) \wedge \neg \textit{isPrimitive}(T_k)$}{
            \If{$\neg cmp_{T_k}(arg_k,\xi(arg_k))$}{
                $\mathcal{D}\leftarrow \mathcal{D}\cup \langle arg_k, arg_i\rangle$
            }
        }
        \If{$\neg cmp_{\tau}(ret, ret')$}{
            $\mathcal{D}\leftarrow \mathcal{D}\cup \langle \tau@\text{return}, arg_i\rangle$
        }
    }
}
\end{algorithm}\DecMargin{1em}

\subsection{Automated Stub Generation for Precise Taint-Analysis}\label{subsec:stub_gen}

We design \sysname{} to automatically generate summary stubs for native methods based on the dependency relations yielded from the mutation-based dependency generation to improve the taint analysis of DroidSafe. A stub in our scenario is a Java method that captures the data dependency between the arguments and the return value of the original native method. The stubs do not provide the complete runtime behavior of native code. Instead, they provide effective abstractions for the static information flow analysis. In the \emph{accurate analysis stubs} of DroidSafe \cite{DBLP:conf/ndss/GordonKPGNR15}, such abstractions are modeled for only Android platform libraries. This section discusses the details of \sysname{}'s automatic stub generation for the native methods for apps and third-party libraries.

Generating a stub for a native method is to use basic operations and predicates compatible with the implementation of DroidSafe to specify how to propagate taints in the native method from the input to the output. For each object, a field \verb|taint| of type \texttt{DSTaintObject} is created by DroidSafe to store the taint information. Therefore, DroidSafe provides two predicates, \textit{addTaint} and \textit{getTaint}$_T$, in the ADI model \cite{DBLP:conf/ndss/GordonKPGNR15} to help
operate on the content of this field; one can specify the propagation rules related to objects with the two predicates. For the primitive-type variables, DroidSafe has built-in taint computation. Thus, the taint propagation can be represented by basic value operations, e.g., an assignment from variable $a$ to $b$ in the stub to mock the dependency of $b$ on $a$.

As mentioned in Section~\ref{subsec:dep_gen}, \sysname{} generates dependency relations between the input and output of a native method. Thus, the automatic stub generation ignores the internal data flows of a native method; instead, it directly builds the data dependency between the input and output (i.e., the arguments and the return value). In all, depending on whether the inputs and outputs are of primitive types ($T_p$) or non-primitive types ($T_{np}$), we design different rules for the four situations as follows:

\begin{compactenum}[1.]
\item  {\bf $T_{np}$ output depends on $T_{np}$ input.} If the type of input is compatible with the type of output, we assign the input to the output; otherwise, we use \textit{getTaint}$_{T_{np}}$ and \textit{addTaint} to adapt the taint of input into the taint of output.

\item  {\bf $T_{np}$ output depends on $T_{p}$ input.} We use \textit{addTaint} to add the value of input itself into the field \verb|taint| of output.

\item {\bf $T_{p}$ output depends on $T_{np}$ input.} We use \textit{getTaint}$_{T_p}$ of the input to return a primitive-type value and assign it to the output.

\item {\bf $T_{p}$ output depends on $T_{p}$ input.} We assign the input to the output after proper type casting of the input.
\end{compactenum}
\noindent When the output depends on multiple primitive-type inputs, we define an operator $\oplus$ over primitive-type value to sum up the value of these inputs and assign the result to the output. When the output depends on multiple non-primitive-type inputs, the summary computations are built up sequentially based on the above rules.

\section{Implementation Issues}\label{sec:implementation}

In this section, we discuss several important issues we should deal with in our implementation.

\paragraph{Implementing the binary analysis} The lightweight static binary analysis in Section~\ref{subsec:ida} can support different ISAs, including ARM (armeabi and armeabi-v7a), ARM-64 (arm64-v8a), X86, and X86-64. The reason we fail to analyze MIPS/MIPS-64 binaries is we do not have an effective method that maps the IDA-disassembled information of a Java-method callsite in native code to the Java method. The applicability of the binary analysis relies on the capability of IDA on control-flow analysis of different ISAs. There are several cases we have to address for this binary analysis.

Our static binary analysis discussed in Section~\ref{subsec:ida} is designed to deal with the cases that sources or sinks are called directly in the native code. Suppose the native code calls another Java method that propagates sensitive data to a sink or takes sensitive data from a source. In that case, the analysis cannot build the correlation between the native method and the sink or the source. For this case, we use \emph{multiple folds of source/sink updating}. We assume, for each app, there are $k$ native methods $\textbf{n}=\{n_1, \ldots, n_k\}$ and $l$ Java methods called on the CFG of the native side of these native methods, i.e. $\textbf{m}=\{m_1,\ldots, m_l\}$. The current set of sources and sinks are respectively \textbf{sc} and \textbf{sk}. Our source/sink updating principle is as follows:
\begin{compactenum}[1.]
\item \emph{Backward sink updating}: For some $n_i\in\textbf{n}$ such that $n_i$ has been inserted to \textbf{sk} according to Section~\ref{subsec:ida}, we take \textbf{m} as the source list and apply our data-flow analysis to detect data flows from each method in \textbf{m} to $n_i$. For any $m_j\in\textbf{m}$ such that a data flow to $n_i$ is discovered (i.e., $m_j\rightsquigarrow n_i$), we update \textbf{sk} with a new sink $m_j$ and start the second fold of static binary analysis in Section~\ref{subsec:ida}, until no new sinks can be discovered.

\item \emph{Forward source updating}: For some $n_i\in\textbf{n}$ such that $n_i$ has been inserted to \textbf{sc} according to Section~\ref{subsec:ida}, we take \textbf{m} as the sink list and apply our data-flow analysis to detect flow from $n_i$ to each method in \textbf{m}. For any $m_j\in\textbf{m}$ such that $n_i\rightsquigarrow m_j$, we update \textbf{sc} with $m_j$ and start the second fold of static binary analysis, until the \textbf{sc} cannot be updated.
\end{compactenum}

\noindent After multiple folds of source/sink updating, \textbf{sc} and \textbf{sk} will eventually reach a fixpoint since the number of methods is limited for any app.

To make the data-flow analysis in each fold precise, the mutation-based analysis and updated stubs should be applied in advance. Meanwhile, the wrapping of the calls to proxy source methods discussed in Section~\ref{subsec:ida}, as a step of stub updating, depends on the result of each fold, which indeed makes the stub updating and static binary analysis mutually recursive. To trade-off between completeness and scalability, we specify a finite depth of folds to control the complexity of source/sink updating in our implementation. As a result, by combining the native-side control-flow correlation and Java-side data-flow analysis, which makes the back-and-forth interactions between native code and Java code trackable, the source/sink updating can discover more native methods as sources and sinks. For instance, when setting the depth of folds as 2, we can add the native method \verb|n_1| in Fig.~3 of \cite{DBLP:conf/ccs/WeiLOCZ18} to the sink list.

\paragraph{Implementing dynamic dependency generation} The dynamic analysis in Algorithm~\ref{algo:dependency_gen} is implemented into two phases. The first phase is a program generator that statically analyzes the signature of native methods of the original app and generates code for Algorithm~\ref{algo:dependency_gen} as a new Android app. In this procedure, we use the Soot framework \cite{DBLP:conf/cascon/Vallee-RaiCGHLS99}. The second phase is running the generated app in an Android emulator to derive the dependency relation $\mathcal{D}$. We use \verb|DexClassLoader| to load the classes and shared objects from the original app and use reflection-based programming to call the native methods. The generated app can support both Android 4.4 and Android 8.0 runtime. For the scalability issue, we limit the parameters used in the dynamic analysis, such as the domain of primitive types, the size of arrays, the length of strings, and the times of mutations (i.e., \textit{BOUND} in Algorithm~\ref{algo:dependency_gen}). Because the dependency relation $\mathcal{D}$ is an under-approximation of the real dependencies and we feed the pairwise execution of the native method with random inputs, larger values for these parameters improve the completeness of data-flow analysis but increase the overall computational cost of the analysis.

Preparing the inputs with non-primitive types for each iteration of Algorithm~\ref{algo:dependency_gen} requires us to build objects with random content for the potential mutations or clones. If the type of argument of the native method is an interface or abstract class, we apply some type inferences to map the argument to some concrete class or subtype to find available object constructors. Specifically, the dependency generation will search for all the non-primitive subtypes implementing the interface or derived from the abstract class, and perform a \textit{clone}, \textit{mutate}, or \textit{cmp} operation on each subtype. On the other hand, in preparing an input argument with primitive type or immutable non-primitive type, e.g., \verb|String|, such an argument is immutable in the pairwise executions of the native method; therefore, we use the same copy instead of cloning it. For reliability, we resort to several libraries \cite{deep-clone, java-util} to implement the \textit{clone}$_T$, \textit{cmp}$_T$, and \textit{mutate}$_T$ operations. These libraries also help deal with the recursive types of arguments and return value. The static field of type $T$ argument of the native method cannot be cloned deeply as did on the heap objects. To mitigate this constraint, we derive a subclass $T' <: T$ and create a shadow static field in $T'$ for each static field of $T$. Then, the instances of $T'$ are used in the dependency generation.

\paragraph{Special native methods with no dependency} There are two forms of native methods for which our dependency generation in Algorithm~\ref{algo:dependency_gen} is not applicable. Instead, we deal with them specially:
\begin{compactenum}[1.]
\item For static native method with no argument, we cannot build up the dependency relation. However, the return values of the pairwise executions may still vary due to the potential indeterminate impact from the native side. We treat this case conservatively as a taint generation.

\item For static native method that has only primitive-type arguments and returns \verb|void|, the derived dependency relation is always empty. We generate an empty stub for such method.
\end{compactenum}


\section{Evaluation}\label{sec:evaluation}

This section investigates the accuracy and effectiveness of \sysname{} by comparing it with DroidSafe \cite{DBLP:conf/ndss/GordonKPGNR15} and JN-SAF \cite{DBLP:conf/ccs/WeiLOCZ18}.

The evaluations involve three datasets of Android apps, as presented in Table~\ref{tab:dataset}. For the accuracy comparison, we use dataset S1, which consists of 23 apps of NativeFlowBench \cite{DBLP:conf/ccs/WeiLOCZ18}, the app in Fig.~\ref{fig:example}, and 119 apps of DroidBench 2.0 \cite{droidbench}, with complete ground truths of the sensitive information flows in them.
Dataset S2 and S3 are used to evaluate the effectiveness of different tools on real-world apps and malware. We summarized the number of shared objects in different ISAs in the apps of S2 and S3. There are 15,203 native libraries in datasets S2 and S3. 73.0\% of them (11,096 shared objects) are in ARM/ARM-64 (armeabi, armeabi-v7a, and arm64-v8a), and 21.1\% of them (3,215 shared objects) are in X86/X86-64. Only around 5.9\% are in MIPS/MIPS-64, which we do not support analyzing. On the other hand, in datasets S2 and S3, we find no \emph{native Activity} component, which we fail to resolve and is also reported to be very rare in the datasets of JN-SAF\cite{DBLP:conf/ccs/WeiLOCZ18}.

Our experiments are conducted on an elastic compute service with 2.5GHz$\times$8 Intel Xeon(Cascade Lake) Platinum 8269CY CPU, 64GB RAM, Linux 4.4.0-174-generic kernel (Ubuntu 16.04). To avoid bias on the results of the information-flow analysis, we use the same source and sink list for each tool. We merge the source/sink lists of DroidSafe and JN-SAF and use the merged source and sink list in our evaluation. The source list contains 7,873 source methods, and the sink list contains 3,551 sink methods.

\begin{table}[!t]
\renewcommand{\arraystretch}{1.3}
\caption{Datasets for the evaluation}\label{tab:dataset}
\centering
\begin{tabular}{c r l}
\hline
Dataset & \#App & Description \\
\hline
S1 & 143 & NativeFlowBench \cite{DBLP:conf/ccs/WeiLOCZ18}, the example in Fig.~\ref{fig:example}, \\
& & and DroidBench 2.0 \cite{droidbench} \\

S2 & 5,096 & Real-world apps with native code. Released  \\
& &  Jul.-Oct. 2019 on Google Play, and got through \\
& & AndroZoo \cite{Allix:2016:ACM:2901739.2903508} \\

S3 &  2,052 & Malware with native code from Drebin \cite{DBLP:conf/ndss/ArpSHGR14}, \\
& & DroidAnalytics \cite{10.1109/TrustCom.2013.25} and CICInvesAndMal2019 \\
& & \cite{8888430} \\
\hline
\end{tabular}
\end{table}

\subsection{Accuracy of \sysname{}}\label{subsec:rq1}

\begin{table*}[!t]
\renewcommand{\arraystretch}{1.3}
\caption{Results on the apps with native code in dataset S1}\label{tab:accuracy}
\centering \scriptsize
\begin{tabular}{l | c | c c c | c c c | c c c}
\hline
App Name & GTP & \multicolumn{3}{c|}{DroidSafe} &  \multicolumn{3}{c|}{JN-SAF} & \multicolumn{3}{c}{\sysname{}} \\
\cline{3-11}
& & TP & FP & FN & TP & FP & FN & TP & FP & FN \\
\hline
native\_source & 1 & 0 & 0 & 1 & 1 & 0 & 0 & 1 & 0 & 0 \\

native\_nosource & 0 & 0 & 0 & 0 & 0 & 0 & 0 & 0 & 0 & 0 \\

native\_source\_clean & 0 & 0 & 0 & 0 & 0 & 0 & 0 & 0 & 0 & 0 \\

native\_leak & 1 & 0 & 0 & 1 & 1 & 0 & 0 & 1 & 0 & 0 \\

native\_leak\_array & 1 & 0 & 0 & 1 & 1 & 0 & 0 & 1 & 0 & 0 \\

native\_leak\_dynamic\_register & 1 & 0 & 0 & 1 & 1 & 0 & 0 & 1 & 0 & 0 \\

native\_dynamic\_register\_multiple & 1 & 0 & 0 & 1 & 1 & 0 & 0 & 1 & 0 & 0 \\

native\_noleak & 0 & 0 & 0 & 0 & 0 & 0 & 0 & 0 & 0 & 0 \\

native\_noleak\_array & 0 & 0 & 0 & 0 & 0 & \cellcolor{lightgray}1 & 0 & 0 & 0 & 0 \\

native\_method\_overloading & 1 & 0 & 0 & 1 & 1 & 0 & 0 & 1 & 0 & 0 \\

native\_multiple\_interactions & 1 & 0 & 0 & 1 & 1 & 0 & 0 & 1 & 0 & 0 \\

native\_multiple\_libraries & 1 & 0 & 0 & 1 & 1 & 0 & 0 & 1 & 0 & 0 \\

native\_complexdata & 1 & 0 & 0 & 1 & 1 & 0 & 0 & 1 & \cellcolor{lightgray}1 & 0 \\

native\_complexdata\_stringop & 0 & 0 & 0 & 0 & 0 & \cellcolor{lightgray}1 & 0 & 0 & 0 & 0 \\

native\_heap\_modify & 1 & 0 & 0 & 1 & 1 & 0 & 0 & 1 & 0 & 0 \\

native\_set\_field\_from\_native & 2 & 0 & 0 & 2 & 2 & 0 & 0 & 2 & 0 & 0 \\

native\_set\_field\_from\_arg & 2 & 0 & 0 & 2 & 2 & 0 & 0 & 2 & 0 & 0 \\

native\_set\_field\_from\_arg\_field & 2 & 0 & 0 & 2 & 2 & 0 & 0 & 2 & 0 & 0 \\
\hline
native\_pure & 1 & 0 & 0 & 1 & 1 & 0 & 0 & 0 & 0 & \cellcolor{lightgray}1 \\

native\_pure\_direct & 1 & 0 & 0 & 1 & 1 & 0 & 0 & 0 & 0 & \cellcolor{lightgray}1 \\

native\_pure\_direct\_customized & 1 & 0 & 0 & 1 & 1 & 0 & 0 & 0 & 0 & \cellcolor{lightgray}1 \\
\hline
icc\_javatonative & 1 & 0 & 0 & 1 & 1 & 0 & 0 & 1 & 0 & 0 \\

icc\_nativetojava & 1 & 0 & 0 & 1 & 1 & 0 & 0 & 1 & 0 & 0 \\
\hline
example\_fig1 & 2 & 1 & 0 & 1 & 1 & 0 & \cellcolor{lightgray}1 & 2 & 0 & 0 \\
\hline
Total  & 23 & 1 & 0 & 22 & 22 & 2 & 1 & 20 & 1 & 3 \\
\hline
\end{tabular}
\end{table*}

\begin{table*}[!t]
\renewcommand{\arraystretch}{1.3}
\caption{Results on the apps of DroidBench 2.0 in dataset S1}\label{tab:accuracy2}
\centering \scriptsize
\begin{tabular}{l | c | c | c c c | c c c | c c c }
\hline
App Category & \#apps & GTP &  \multicolumn{3}{c|}{DroidSafe} &  \multicolumn{3}{c|}{JN-SAF} & \multicolumn{3}{c}{\sysname{}} \\
\cline{4-12}
& & & TP & FP & FN & TP & FP & FN & TP & FP & FN \\
\hline
Aliasing & 1 & 0 & 0 & 0 & 0 & 0 & 1 & 0 & 0 & 0 & 0 \\

Arrays and Lists & 7 & 3 & 3 & 4 & 0 & 0 & 4 & 3 & 3 & 4 & 0 \\

Callbacks & 15 & 17 & 17 & 6 & 0 & 10 & 3 & 7 & 17 & 6 & 0 \\

Field and Obj Sens & 7 & 2 & 2 & 2 & 0 & 2 & 0 & 0 & 2 & 2 & 0 \\

IAC & 3 & 8 & 8 & 12 & 0 & 4 & 0 & 4 & 8 & 12 & 0 \\

ICC & 18 & 24 & 24 & 5 & 0 & 17 & 0 & 7 & 24 & 5 & 0 \\

Lifecycle & 17 & 17 & 17 & 4 & 0 & 10 & 0 & 7 & 17 & 4 & 0 \\

General Java & 23 & 20 & 20 & 8 & 0 & 11 & 3 & 9 & 20 & 8 & 0 \\

Misc Android-Specific & 12 & 11 & 9 & 2 & 2 & 7 & 0 & 4 & 9 & 2 & 2 \\

Implicit Flows & 4 & 8 & 2 & 0 & 6 & 0 & 0 & 8 & 2 & 0 & 6 \\

Reflection & 4 & 4 & 4 & 0 & 0 & 1 & 0 & 3 & 4 & 0 & 0 \\

Threading & 5 & 5 & 4 & 1 & 1 & 4 & 0 & 1 & 4 & 1 & 1 \\

Emulator Detection & 3 & 6 & 0 & 0 & 6 & 4 & 0 & 2 & 0 & 0 & 6 \\
\hline
Total & 119 & 125 & 110 & 44 & 15 & 70 & 11 & 55 & 110 & 44 & 15 \\
\hline
\end{tabular}
\end{table*}

On dataset S1, we evaluate the accuracy of \sysname{} compared with JN-SAF and the original DroidSafe. We separate the evaluation into two parts. In Table~\ref{tab:accuracy}, we present the results on the test cases containing native code. In Table~\ref{tab:accuracy2}, we show the results on DroidBench 2.0, a general benchmark suite to evaluate the accuracy of different taint analyses for Android apps. Note that none of the apps in DroidBench 2.0 contains native code. In Table~\ref{tab:accuracy} and Table~\ref{tab:accuracy2}, we present the \emph{true positives} (TP), \emph{false positives} (FP), and \emph{false negatives} (FN) of each approach. GTP stands for the \emph{ground-truth positives}. We summarize the measurements of different approaches in Table~\ref{tab:metrics}. \sysname{} can reach higher recall and F1-score than JN-SAF. Therefore, we conclude that JN-SAF may be preferred in circumstances where high precision (less FPs) is more desired than high recall (less FNs), e.g., exploit construction, while \sysname{} may be better when high recall is more favored, e.g., information-leakage defense. Thus, we use the F1-score to measure the overall efficacy for an unbiased comparison. We next discuss in detail the false positives and false negatives we observed in our experiment.

\paragraph{False positives of native-code-involved flows}
Since DroidSafe can only build empty stubs for native methods introduced by the app itself, the potential taint propagation will be cut off. As a result, DroidSafe cannot report any positive for NativeFlowBench at all.
JN-SAF detects two false positives for NativeFlowBench.
One of them is reported in the analysis of \verb|native_noleak_array|; in this case, JN-SAF cannot distinguish different indices of array elements; therefore, it treats the tainting on one element to be propagated over the whole array. The other one is reported for \verb|native_complexdata_stringop|; in this case, JN-SAF cannot precisely analyze the string operations; thus, it conservatively treats more fields of data structures as tainted fields. For both cases, the dependency generation of \sysname{} can successfully figure out more accurate dependencies.
As for the false positives of \sysname{}, in the analysis of \verb|native_complexdata|, both native methods call the sink method \verb|LOGI|, but only one \verb|LOGI| releases sensitive data. Because \sysname{} conservatively takes both native methods as sinks due to their control-flow correlation with \verb|LOGI| and does not consider the sensitivity of data delivered into \verb|LOGI|, \sysname{} then reaches a false positive.

\paragraph{False negatives of native-code-involved flows}
As we explained earlier, DroidSafe is incapable of detecting any sensitive flows in NativeFlowBench. Thus, all the ground-truth flows are false negatives of DroidSafe, including the native-code-involved flow in our motivating example in Fig.~\ref{fig:example}.
Meanwhile, JN-SAF also misses the native-code-involved flow in Fig.~\ref{fig:example}. Our investigation indicates that JN-SAF's implementation issue makes its heap manipulation summary imprecise on this example. In contrast, \sysname{}'s mutation-based dynamic analysis figures out better summaries for more complicated memory accesses.
As a comparison, the false negatives reported by \sysname{} are all due to its inability to deal with the native Activity components, as demonstrated by apps \texttt{native\_pure}, \texttt{native\_pure\_direct}, and \texttt{native\_pure\_direct\_customized} in Table~\ref{tab:accuracy}.

\begin{table}[t]
\renewcommand{\arraystretch}{1.3}
  \caption{Metrics comparison of different tools on dataset S1}
  \label{tab:metrics}
  \centering
  \begin{tabular}{l|r r r}
\hline
    Tool & Precision(\%) & Recall(\%) & F1(\%) \\
\hline
    DroidSafe & 71.6 & 75.0 & 73.3 \\
    JN-SAF & 87.6 & 62.2 & 72.7 \\
    \sysname{} & 74.3 & 87.8 & 80.5 \\
\hline
  \end{tabular}
\end{table}




\paragraph{Differences in detecting native-code-uninvolved flows}
Since none of the apps in Table~\ref{tab:accuracy2} contains native code and \sysname{} is built upon DroidSafe, the two tools report identical results. Next, we compare \sysname{} with JN-SAF.
According to Table~\ref{tab:accuracy2}, we observe that \sysname{} and JN-SAF have diverse behaviors. In most app categories of DroidBench, \sysname{} reports both more true positives and false positives than JN-SAF, while JN-SAF reports more false negatives. We infer the reason for such difference is that JN-SAF's summary-based bottom-up data-flow analysis is specialized in the inter-language data-flow analysis but is limited in propagating the points-to information to the callee. In other words, we believe \sysname{} achieves better context sensitivity than JN-SAF. Therefore, JN-SAF cannot achieve similar effectiveness compared with the state-of-the-art context-, object- and field-sensitive interprocedural data-flow analysis, e.g., \cite{DBLP:conf/ndss/GordonKPGNR15, DBLP:conf/pldi/ArztRFBBKTOM14}. Besides, merging the detection results of DroidSafe and JN-SAF to achieve better coverage is not trivial since there may be contradictory detection results from them on native-code-involved flows. In contrast, \sysname{} is integrated into DroidSafe, which prevents this possibility.

\subsection{Capability of \sysname{} compared with another dynamic taint analysis}

\sysname{} has a dynamic phase to generate the data dependencies for native methods, indicating that the incompleteness of dependencies derived by our dynamic analysis is a source of false negatives. Therefore, we evaluate the code coverage of native code during our mutation-based dependency generation. We use static binary instrumentation with Dyninst \cite{dyninst} to profile the basic blocks of the shared object files reached by our dynamic analysis. The code coverage of the native code is measured by
\begin{displaymath}
\frac{\textit{\#reached basic blocks}}{\textit{\#basic blocks}}\times 100\%
\end{displaymath}
For the test cases with native code in dataset S1, each native method has 6.7 basic blocks on average. We show the code coverage in Table~\ref{tab:coverage}. Besides the 11 apps of NativeFlowBench reporting instrumentation I/O errors and 119 apps of DroidBench containing no native code, the code coverage on the rest of the apps is 53.1\% on average (29.2\%$\sim$100\%) under the default configuration of \sysname{}.  Several cases have low coverage rates and certain control-flow branches are missed, but the branch misses have not triggered any false negative in the later static taint analysis of \sysname{}.

To compare the capability of taint analysis with the state-of-the-art dynamic taint analysis systems, e.g., NDroid \cite{DBLP:conf/dsn/QianLSC14, DBLP:journals/tifs/XueQZLZSC19}, we label the test cases in dataset S1 with the scenarios (\textit{P1}$\sim$\textit{P6}) of information leakages defined in Fig.~3 of \cite{DBLP:journals/tifs/XueQZLZSC19}.
Each scenario depicts one pattern of whether and how information flow crosses the native and Java contexts. For example, P2 represents the scenario where sensitive flows start from the Java context, then switch to the native context, and eventually return to the Java context. Our labeling results of test cases are listed in Table~\ref{tab:coverage}.
Based on our labeling, considering the TP/FP/FN presented in Table~\ref{tab:accuracy} and Table~\ref{tab:accuracy2}, we can see that \sysname{} is able to reach high precision and high recall for test cases of scenario \textit{P1}$\sim$\textit{P5}. The only exception is scenario \textit{P6}, which represents the case that the sensitive flow stays at the native side. For such native-only sensitive flows, if they are implemented by native Activity, \sysname{} will report only false negative, as demonstrated by the three native Activity cases in Table~\ref{tab:accuracy}. Otherwise, our static binary analysis will deem the native methods containing such flows as both sources and sinks, but it cannot tell the exact locations of the sources or sinks in the native method.
Because we failed to adjust the sources and sinks of NDroid with the source/sink list of \sysname{}, we are unable to give a fair comparison of the metrics with ground truths.

\begin{table}[!t]
\renewcommand{\arraystretch}{1.3}
\caption{Information Leakage Scenarios of dataset S1 and code coverage of native code}\label{tab:coverage}
\centering \scriptsize
\begin{tabular}{l | c | c }
\hline
App Name & Scenario & Coverage(\%) \\
\hline
native\_source & \textit{P3} & 57.1 \\

native\_nosource & No leak & 100.0 \\

native\_source\_clean & No leak & 100.0 \\

native\_leak & \textit{P1} & N/A \\

native\_leak\_array & \textit{P1} & N/A \\

native\_leak\_dynamic\_register & \textit{P1} & N/A \\

native\_dynamic\_register\_multiple & \textit{P1} & 62.5 \\

native\_noleak & No leak & 100.0 \\

native\_noleak\_array & No leak & N/A \\

native\_method\_overloading & \textit{P1} & N/A \\

native\_multiple\_interactions & \textit{P1}+\textit{P4} & N/A \\

native\_multiple\_libraries & \textit{P1} & N/A \\

native\_complexdata & \textit{P1} & 70.0 \\

native\_complexdata\_stringop & \textit{P1} & N/A \\

native\_heap\_modify & \textit{P3} & 68.0 \\

native\_set\_field\_from\_native & \textit{P3} & 48.7 \\

native\_set\_field\_from\_arg & \textit{P2} & 100.0 \\

native\_set\_field\_from\_arg\_field &  \textit{P2} & 100.0 \\

native\_pure* (3 apps) & \textit{P6} & N/A \\



icc\_javatonative & \textit{P1} & 29.2 \\

icc\_nativetojava & \textit{P2} & 80.6 \\

example\_fig1 & \textit{P2}+\textit{P5} & 100.0 \\
\hline
DroidBench (119 apps) & \textit{P5}/No leak & No \textit{.so} \\
\hline
Avg. & -- & 53.1 \\
\hline
\end{tabular}
\end{table}

\subsection{Improvement of \sysname{} upon DroidSafe}

\sysname{} is built upon the backend of DroidSafe and its ADI model. Based on the precision and recall metrics presented in Section~\ref{subsec:rq1}, we evaluate how \sysname{} can improve DroidSafe by introducing native code analysis and new stub generation. Furthermore, we apply \sysname{} to understand how shared object files are involved in sensitive information flows in real-world apps and malware.

Our experiments are conducted on dataset S2 and S3. The number of new sensitive flows introduced by native code is considerable, as illustrated in the first three columns of Table~\ref{tab:num-flow}. For the app in S2, DroidSafe can detect 5,052 sensitive flows while \sysname{} can detect 6,427, with an increase rate of 27.2\%. For the app in S3, DroidSafe detects 14,330 sensitive flows, and \sysname{} can detect 16,920, with an increase rate of 18.1\%. We are aware that the incremental component contains false positives. Therefore, we hope to evaluate the true positive increase rates. However, due to the difficulty in obtaining the ground truths of real-world apps, we can only use the precision yielded on S1 to estimate the increase rates of true positives on S2 and S3. In detail, with the precision of DroidSafe being 71.6\%, we estimate that DroidSafe detects 3,617 and 10,260 true positives for S2 and S3, respectively. In comparison, we estimate that \sysname{} detects 4,775 and 12,571 true positives, with increase rates being 32.0\% and 22.5\% for S2 and S3, respectively. In fact, we have considered relying on runtime information to collect ground truth for dataset S2 and S3, which turned out to be challenging. On one hand, determining if a profiled execution trace carries sensitive data-flows incurs manual efforts, which is unscalable, while relying on any automated tool would introduce inaccuracy to the ground truth. On the other hand, the collected ground truth is likely to be incomplete, which reduces the credibility of the ground truth. Therefore, we leave the ground-truth collection from runtime information as future work.

To figure out the characteristics of sensitive flows caused by native code, we classify and rank the sensitive flows in the apps of S2 and S3 detected by \sysname{} and compare with the respective numbers of flows detected by DroidSafe. The top-10 categories of sensitive data-flows detected by \sysname{} are listed in Table~\ref{tab:ranking-droidsafe}. We know the most common sensitive flows are in the category \verb|IO|$\rightsquigarrow$\verb|IO|, no matter whether we consider the behavior of native code. An example is reading the content of a file and writing it to another file. There are three categories of sensitive flows increasing enormously after we consider analyzing the native code, i.e., \verb|FILE_INFORMATION|$\rightsquigarrow$\verb|IO|, \verb|LOCATION|$\rightsquigarrow$\verb|NETWORK|, and \verb|UNIQUE_IDENTIFIER|$\rightsquigarrow$\verb|NETWORK|. We investigate the reason for such increases. We found that the increase in \verb|FILE_INFORMATION|$\rightsquigarrow$\verb|IO| flows highly depends on the native libraries of information pushing (libjcore.so, libcocklogic.so, libbdpush.so), map/location (libBaiduMapSDK.so, libtencentloc.so, liblocSDK.so), image processing (libgifimage.so, libwebpbackport.so), and exception report (libBugly.so, libBugtags.so). The increase in \verb|LOCATION|$\rightsquigarrow$\verb|NETWORK| flows frequently depends on the native library of SQLite storage (libsqlc-native-driver.so). The increase in \verb|UNIQUE_IDENTIFIER|$\rightsquigarrow$\verb|NETWORK| flows depends on the native library of privilege promotion (libandroidterm.so). Also, some unknown native libraries, e.g., libopenterm.so, are frequently involved in the apps with these sensitive flows.

\begin{table}[t]
\renewcommand{\arraystretch}{1.3}
  \caption{Number of sensitive flows detected by different approaches}
  \label{tab:num-flow}
  \centering
  \begin{tabular}{c|r r | r r | r r}
\hline
  & \multicolumn{2}{c|}{\#flows (DroidSafe)} & \multicolumn{2}{c|}{\#flows (\sysname{})} & \multicolumn{2}{c}{\#flows (JN-SAF)} \\
  Dataset & Total & Avg. & Total & Avg. & Total & Avg. \\
\hline
    S2 &  5,052 &  8.28 & 6,427 & 10.50 & 497 & 2.54 \\
    S3 & 14,330 & 94.28 &  16,920 &  111.32 & 635 & 4.44 \\
\hline
  \end{tabular}
\end{table}

\begin{table}[t]
\renewcommand{\arraystretch}{1.3}
  \caption{Top ranking sensitive flows detected by \sysname{} compared with DroidSafe}
  \label{tab:ranking-droidsafe}
  \centering\scriptsize
  \begin{tabular}{c r r r}
\hline
source$\rightsquigarrow$sink & \#flows & \#flows & Increasing rate \\
 & (DroidSafe) & (\sysname{}) & \\
\hline
\verb|IO|$\rightsquigarrow$\verb|IO| & 2,128 & 2,209 & 3.81\% \\
\verb|FILE_INFORMATION|$\rightsquigarrow$\verb|IO| & 486 & 1,608 & 230.86\% \\
\verb|NETWORK|$\rightsquigarrow$\verb|NETWORK| & 920 &1,040  & 13.04\% \\
\verb|IO|$\rightsquigarrow$\verb|NETWORK| & 906 &947 & 4.53\% \\
\verb|LOCATION|$\rightsquigarrow$\verb|NETWORK| & 552 &845 & 53.08\% \\
\verb|UNIQUE_IDENTIFIER|$\rightsquigarrow$\verb|NETWORK| & 531 & 718 & 35.22\% \\
\verb|NETWORK|$\rightsquigarrow$\verb|IO| & 572 & 659 & 15.21\% \\
\verb|content.res|$\rightsquigarrow$\verb|IO| & 563 & 563 & 0\% \\
\verb|FILE_INFORMATION|$\rightsquigarrow$\verb|NETWORK| & 495 & 555 & 12.12\% \\
\verb|IO|$\rightsquigarrow$\verb|IPC| & 516 & 521 & 0.97\% \\
\hline
  \end{tabular}
\end{table}

\subsection{Differences between \sysname{} and JN-SAF when applied to real-world apps}

\begin{table*}[!ht]
\renewcommand{\arraystretch}{1.3}
  \caption{Detailed results of the sensitive flows detected by both JN-SAF and \sysname{} (Notations: JS=JN-SAF, $\mu$D = \sysname{}, FD=FlowDroid)}
  \label{tab:detail-jnsaf}
  \centering\scriptsize
  \begin{tabular}{c|c | c | c | c | c | c | c}
\hline
Category & \#apps & \textit{\#flows}$_{\text{JS}}$ & \textit{flows}$_{\text{JS}} \cap$ \textit{flows}$_{\mu\text{D}}$ & \textit{flows}$_\text{JS} - $\textit{flows}$_{\mu\text{D}}$ & \textit{\#flows}$_{\mu\text{D}}$ & \textit{\#flows}$_{\text{FD}}$ & \textit{\#flows}$_{\mu\text{D}\cap \text{FD}}$\\
(Dataset) & ($\mu$D \& JS) & & & & & & \\ \hline

 N/A (S2) & 2 & 2 & \verb|USER_INPUT|$\rightsquigarrow$\verb|IPC|:1 & \verb|USER_INPUT|$\rightsquigarrow$& 43 & 44 & 12.5 \\
 & & & & \verb|SHARED_PREFERENCES|:1 & & & \\ \hline

 N/A (S2) & 1 & 1 & $\emptyset$ & \verb|IPC|$\rightsquigarrow$\verb|LOG|:1 & 108 & 37 & 16 \\ \hline

 N/A (S2) & 1 & 1 & $\emptyset$ & \verb|USER_INPUT|$\rightsquigarrow$ & 23 & 46 & 8 \\
 & & & & \verb|SHARED_PREFERENCES|:1 & & & \\\hline

DroidKungFu & 7 & 1 & \verb|IPC|$\rightsquigarrow$\verb|IPC|:1 & $\emptyset$ &  380 & 105.6 & 13.3 \\
(Drebin) & & & & & & & \\
\hline

Xsider & 2 & 4 & \verb|UNIQUE_IDENTIFIER|$\rightsquigarrow$\verb|LOG|:2,  & $\emptyset$ &  46 & 89 & 26.5 \\
(Drebin) & & & \verb|IPC|$\rightsquigarrow$\verb|LOG|:2 & & & & \\\cline{2-8}
& 1 & 2 & \verb|UNIQUE_IDENTIFIER|$\rightsquigarrow$\verb|LOG|:2 & $\emptyset$  & 41 & 105 & 31 \\ \cline{2-8}
& 1 & 2 & \verb|UNIQUE_IDENTIFIER|$\rightsquigarrow$\verb|LOG|:2 & $\emptyset$ & 35 & 99 & 32 \\ \hline

Mobinauten & 1 & 20 & \verb|UNIQUE_IDENTIFIER|$\rightsquigarrow$\verb|LOG|:2 & \verb|IPC|$\rightsquigarrow$\verb|IPC|:15, & 47 & 105 & 18 \\
(Drebin) & & &  & \verb|UNIQUE_IDENTIFIER|$\rightsquigarrow$ & & & \\
& & &  & \verb|SHARED_PREFERENCES|:3 & & & \\ \hline

Adrd & 1 & 1 & $\emptyset$ & \verb|os|$\rightsquigarrow$\verb|IPC|:1 & 64 & 24 & 2 \\
(Drebin) & & & & & & & \\ \hline

KungFu & 3 & 1 & \verb|IPC|$\rightsquigarrow$\verb|IPC|:1 & $\emptyset$ & 380 & 97 & 14.3 \\
(DroidAnalytics) & & & & & & & \\
 \hline

faketaoBao & 1 & 1 & $\emptyset$ & \verb|os|$\rightsquigarrow$\verb|IPC|:1 & 36 & 11 & 1 \\
(CICInvesAndMal2019) & & & & & & & \\
 \hline
  \end{tabular}
\end{table*}

In our experiment, we observed that \sysname{} detected a lot more sensitive flows than JN-SAF did on datasets S2 and S3. As shown in the last column of Table~\ref{tab:num-flow}, JN-SAF detected 497 sensitive flows on S2 and 635 sensitive flows on S3, while \sysname{} detected 6,427 and 16,920 sensitive flows on S2 and S3, respectively. Thus, in this section, our goal is to interpret such a vast discrepancy. However, due to the lack of ground truths of sensitive information flow in real-world apps, we have to resort to a less decisive approach to performing the comparison. We first attempt to compare false positives; however, we are not aware of a good way to investigate. Instead, we first enumerate the differences we observed as an objective comparison. Then we estimate the potential true positives of \sysname{}.

There are only 21 apps (4 real-world apps in S2 and 17 malware in S3) detected to be vulnerable by both JN-SAF and \sysname{}. In these apps, \sysname{} detected 26 out of 50 sensitive flows that JN-SAF detected, and \sysname{} also detected many more sensitive flows undetectable by JN-SAF. Thus, albeit \sysname{} and JN-SAF give similar results over the NativeFlowBench suite, for real-world apps, we conclude that the two tools demonstrate significant divergences. More details are shown in Table~\ref{tab:detail-jnsaf}. In terms of notation, \textit{flows}$_\text{JS}$ and \textit{flows}$_{\mu\text{D}}$ respectively represent the type of sensitive flows detected by JN-SAF and \sysname{}, in the form of \textit{source\_category}$\rightsquigarrow$\textit{sink\_category}:\#\textit{flows}. For example, in the category \textit{DroidKungFu} of dataset Drebin \cite{DBLP:conf/ndss/ArpSHGR14} (i.e., the 5th row in Table~\ref{tab:detail-jnsaf}), JN-SAF can detect one sensitive flow in each of the seven apps, whose type is \verb|IPC|$\rightsquigarrow$\verb|IPC|. This sensitive flow is also detected by \sysname{} (i.e., \textit{flows}$_\text{JN-SAF} - $\textit{flows}$_{\mu\text{Dep}}=\emptyset$). Meanwhile, \sysname{} can detect other 379 sensitive flows distributed in different flow categories in each of these apps. There are also test cases that most of the sensitive flows detected by JN-SAF are missed by \sysname{}, e.g., the 18 flows in Mobinauten of Drebin.

To investigate the true positives of \sysname{} in more detail, based on our evaluation on dataset S1, for apps without native code (e.g., apps in DroidBench 2.0), \sysname{} built upon DroidSafe is prone to achieve a more complete detection than JN-SAF. It shows that compared to JN-SAF, \sysname{} is more capable of discovering sensitive information flows where native code is not involved. For the real-world apps, we argue that there must exist a large portion of such sensitive information flows. So, for this portion of sensitive information flows, we believe \sysname{} should detect more true positives. Moreover, due to the lack of ground truths, we use another independent taint analyzer, FlowDroid \cite{DBLP:conf/pldi/ArztRFBBKTOM14} (v2.8), to estimate the magnitude of true positives detected by \sysname{}. In Table~\ref{tab:detail-jnsaf}, \textit{\#flows}$_\text{FD}$ is the average number of flows detected in each app by FlowDroid. \textit{\#flows}$_{\mu\text{D}\cap\text{FD}}$ is the average number of flows in each app detected by both FlowDroid and \sysname{}. Because of the difference in the approaches and configurations, we cannot infer the number of true positives outside \textit{flows}$_{\mu\text{D}\cap\text{FD}}$. However, we can deduce the flows in \textit{flows}$_{\mu\text{D}\cap\text{FD}}$ tend to be true positives, which are in general more than the flows detected by JN-SAF. Besides, FlowDroid only uses mock stubs for system-defined native code and should miss sensitive flows related to the native side, which justifies the true positives should be more than \textit{\#flows}$_{\mu\text{D}\cap\text{FD}}$.

\subsection{Analysis efficiency of \sysname{}}

We evaluate the efficiency of our approach on the apps analyzable by both JN-SAF and \sysname{}. The average analysis runtime and the peak memory consumption of both approaches are presented in Table~\ref{tab:efficiency}. For JN-SAF, the analysis is divided into two phases, summary-based bottom-up data-flow analysis (SBDA) and native code analysis (Native). For \sysname{}, our analysis consists of the control-flow based static binary analysis (CFBA), the mutation-based dependency generation (DepGen), the stub generation (StubGen), and the information flow analysis using DroidSafe (DroidSafe). For both approaches, the data-flow analysis is in the lead of the computational cost in general. Because of the difference in the points-to analysis, the SBDA of JN-SAF is generally more efficient than the data-flow analysis of \sysname{} (based on DroidSafe) on both time and memory cost. For the native-side static analysis, \sysname{} resorts to control-flow analysis, while JN-SAF implements an annotation-based data-flow analysis. Besides, the automated stub generation is efficient, and the computational cost can be ignored.

\begin{table}[t]
\renewcommand{\arraystretch}{1.3}
  \caption{Efficiency comparison of JN-SAF and \sysname{} (average on each app)}
  \label{tab:efficiency}
  \centering
  \begin{tabular}{c | c | c | c | c | c}
\hline
 & \multicolumn{4}{c|}{Time(s)} & Mem(GB) \\
 \hline
JN-SAF & \multicolumn{3}{c|}{SBDA} & Native &  \\ \cline{2-5}
 & \multicolumn{3}{c|}{564.50} & 16.67 & 8.44 \\
\hline
\sysname{} & DepGen & StubGen & DroidSafe & CFBA & \\\cline{2-5}
 & 23.20 & 0.00 & 678.96 & 0.89 & 21.56 \\
\hline
  \end{tabular}
\end{table}

\begin{figure}[t]
\centering
\includegraphics[width=\linewidth]{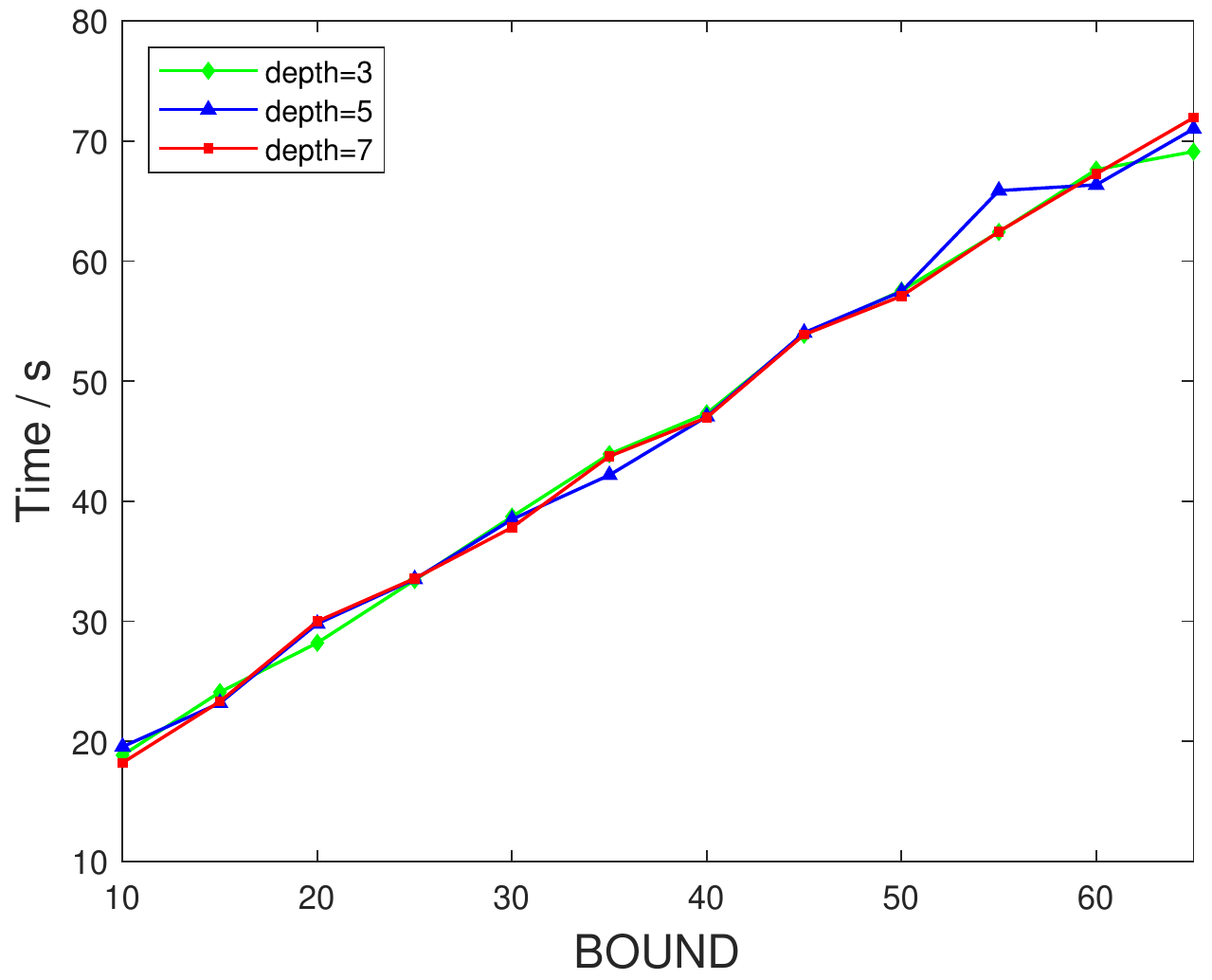}
\caption{Effect of Configurations of Dependency Generation on the Time Cost}
\label{fig:conf-dep-gen}
\end{figure}

The dynamic dependency generation of our approach relies on several configurations. To collect data for the efficiency comparison, we used the default configuration, i.e., \textit{BOUND}=15, and the depth of field applied on the atomic predicates, e.g., $\textit{cmp}_T$ and $\textit{mutate}_T$, is 5. The effect of different \textit{BOUND} values and depths of field on the time costs of dependency generation is illustrated in Fig.~\ref{fig:conf-dep-gen}. The time cost increases linearly with the increment of \textit{BOUND}, for each choice on the depth of field. On the other hand, there is no apparent differentiation in the time costs given different depths of field, indicating that, in our implementation, the argument preparations and comparisons using the atomic predicates are efficient.

\section{Discussion}\label{sec:discuss}

In this section, we discuss the sources of false positives and false negatives introduced by the dynamic analysis performed on the native code (\textit{.so} files) and the threats to the validity of our approach.

\paragraph{Sources of false negatives}
First, \sysname{} cannot deal with native Activity components, thus will not report the sensitive flows inside these components, which lead to false negatives. Second, the randomly constructed values, objects, and object fields that we feed to the pairwise execution of the native method in Algorithm~\ref{algo:dependency_gen} may be insufficient to discover all data dependencies between arguments and return values of native methods, especially dependencies relying on specific system states, e.g., data from other threads. \sysname{} relies on increasing the mutation iterations, i.e., the value of \texttt{BOUND} in Algorithm~\ref{algo:dependency_gen}, to improve the coverage of dynamic analysis and to reduce the chance of such false negatives.

\paragraph{Sources of false positives}
Randomness at the native side can be a factor in causing false positives. For example, if a random event can bypass the input cloning and trigger some differentiation of the output of the native function, the dependency generation algorithm will mistakenly build a dependency between the current mutated input and such output. Another source of false positives is
the binary-level static control-flow analysis that correlates existing sinks with newly added sinks. Such an analysis introduces over-approximations, which may bring spurious correlations of native sinks to the taint analysis, as shown in \texttt{native\_complexdata} in Table~\ref{tab:accuracy}.
Moreover, if the native code accesses a stateful function, such a function might be another source of randomness to the local state of the native code, resulting in false positives.

\paragraph{Other threats}
\sysname{} uses dynamic analysis to generate data dependencies for the native methods and then builds the code stubs based on these data dependencies for the static taint analysis. Such \emph{dynamic-over-static} strategy confronts the difficulty of analyzing obfuscated or hardened apps. Such limitations are common in most of the static taint analyses of Android apps, e.g., \cite{DBLP:conf/ccs/WeiLOCZ18, DBLP:conf/ndss/GordonKPGNR15, DBLP:conf/pldi/ArztRFBBKTOM14}. We expect unpackers, e.g., PackerGrind \cite{DBLP:conf/icse/XueLYWW17, packergrind2}, and deobfuscators, such as \cite{simply}, may help mitigate such limitations. In contrast, the dynamic taint analyses, e.g., \cite{DBLP:conf/osdi/EnckGCCJMS10, DBLP:conf/dsn/QianLSC14}, suffer less from such limitations. On the other hand, our static taint analysis framework inherited from DroidSafe, whose ADI model modified by \sysname{}, mainly supports Android 4.4 APIs, even though our dynamic analysis can support Android 8.0 runtime. Another issue is the difference between the patterns of randomly generated inputs and the specific patterns of the inputs used by app developers. Our random inputs and mutations may discover a considerable number of valid but rarely used data dependencies. An expected improvement is to profile the context of each native method call to capture the commonly used patterns of arguments.


\section{Related Work}\label{sec:related-work}

\paragraph{Security on Native Code of Android} Native code dramatically benefits the performance-critical applications, e.g., games and graphical acceleration, and for the purpose of anti-reverse engineering \cite{DBLP:conf/trustbus/ProtsenkoM15}\cite{DBLP:conf/cisc/WangLZWHLG17}. Using native code and libraries is very popular in mobile scenarios \cite{DBLP:conf/wisec/SunT14}\cite{DBLP:conf/ndss/AfonsoGBFKVDP16}. However, a large portion of native code used in Android apps is migrated from open source projects \cite{DBLP:conf/cisc/WangLZWHLG17}. Meanwhile, the lack of control mechanisms for the execution of native code, and the misuse of domain-specific native functions, pose a significant threat to the security of the Android platform and apps.

The native code and libraries have been treated as an important source of root exploits and root privilege escalation. Fedler et al.\cite{DBLP:conf/ccs/FedlerKS13} propose to control the execution of native binaries and libraries at the system level by modifying \verb|chmod| or customizing specific Java library APIs. RiskRanker \cite{DBLP:conf/mobisys/GraceZZZJ12} compares the native code with the signatures of known root exploits. It also detects whether the encrypted native exploit code is stored in an irregular directory and decrypted for execution at runtime. DroidRanger \cite{DBLP:conf/ndss/ZhouWZJ12} uses native code as features to identify the family of zero-day malicious apps. It relies on a dynamic execution monitor to inspect the runtime behaviors of untrusted code, especially to collect the system calls made by native code. The PREC framework \cite{DBLP:conf/codaspy/HoDGE14} bridges offline behavior learning and runtime anomaly detection to mitigate root exploits. It uses thread-based dynamic analysis to identify system calls originating from risky third-party native code. By matching the client-side runtime system call sequences to the normal behavior model of the app, PREC identifies malicious system call sequences and suppresses the malicious activity in the native thread.

Moreover, native code is usually exploited to dynamically load external malicious code \cite{DBLP:conf/ndss/PoeplauFBKV14} or violate privacy and safety at the application level \cite{DBLP:conf/wisec/SunT14}. User-level sandboxing is a promising approach to compartmentalize the actions of native code and restrict the communication between native code and Java code. NativeGuard \cite{DBLP:conf/wisec/SunT14} confines the potential malicious behaviors of third-party native libraries by separating the native libraries from Android application to another stand-alone application, where native code resides in a different address space and is deprived of unnecessary privileges, to improve the overall security of the application. AppCage \cite{DBLP:conf/ccs/ZhouPWWJ15} proposes a lightweight inner-process native sandbox to prevent the native libraries of an app from modifying data and code outside the sandbox. The native sandbox relies on SFI and is implemented through binary rewriting and instrumentation. NaClDroid \cite{DBLP:conf/esorics/AthanasopoulosK16} also takes a thread-level SFI to confine the untrusted native code. It uses Google's Native Client program in a separated thread to redirect specific calls for loading the modules which host the native code. NativeProtector \cite{DBLP:conf/sec/HongWY16} follows the process-based isolations of NativeGuard and intercepts sensitive native calls to perform fine-grained access control.

Several system-level approaches treat the Java and native code analyses indiscriminately by capturing the critical runtime features, e.g., system calls, of malicious behaviors. Emulator-level debug tools, e.g., \textit{ltrace}/\textit{strace} \cite{DBLP:journals/ijisec/SpreitzenbarthS15, DBLP:conf/ndss/AfonsoGBFKVDP16} and the interception on specific instruction \cite{DBLP:conf/ndss/TamKFC15}, are useful to capture the sensitive system calls, which are drastically relied on by the dynamic analysis for identifying the behaviors of the native part of apps. DroidScope \cite{DBLP:conf/uss/YanY12} is a general-purpose emulation-based framework using simultaneous two-level VM introspection to rebuild the semantics of Java and native components. CrowDroid \cite{DBLP:conf/ccs/BurgueraZN11} uses a client-side app to monitor system calls reflecting the behavior of other apps, collects such system calls on the server side, and builds normality model to detect anomalies of malicious apps. Mobile-Sandbox \cite{DBLP:journals/ijisec/SpreitzenbarthS15} traces the system calls made into shared objects and logs such events as features for the learning-based malicious behavior detection. CopperDroid \cite{DBLP:conf/ndss/TamKFC15} also employs VM introspection to record the system calls regardless of whether they are from Dalvik or native code, and then rebuilds the high-level semantics of objects and behaviors. Afonso et al.\cite{DBLP:conf/ndss/AfonsoGBFKVDP16} implement a dynamic analysis by instrumenting the core libraries of the emulator to monitor the native-side events. The dynamic analysis automatically generates security policies, i.e., white-list of normal system calls and Java methods invoked by native code, for the existing native sandboxing approach, e.g., NativeGuard \cite{DBLP:conf/wisec/SunT14}. The objective of \cite{DBLP:conf/ndss/AfonsoGBFKVDP16} is to derive proper sandboxing policies for native code that can avoid malicious behaviors but facilitate the normal execution of the native code, while the dynamic analysis of \sysname{} is to assist the static taint analysis. Harvester \cite{DBLP:conf/ndss/RasthoferAMB16} combines backward slicing with dynamic code execution to resolve reflective method calls and extract malicious features hidden by reflections. The derived reflection information has been used to improve both static and dynamic taint analysis \cite{DBLP:conf/osdi/EnckGCCJMS10, DBLP:conf/pldi/ArztRFBBKTOM14}. The reflections derived by Harvest are different from the data dependencies derived by our dynamic dependency generation, which makes Harvester and \sysname{} support diverse aspects of static taint analysis.

\paragraph{Impreciseness of Information Flow Analysis for Android} The permission system of Android has been proved in practice insufficient to detect inconspicuous misbehavior, which violates information flow security policy. For example, suppose the data of one component are permitted to be accessed by another component. In that case, this component will have full authority to dispose of the data, including misoperating on or leaking the data without control. More sophisticated taint tracking approaches or information flow analysis are obligatory to avoid this kind of misbehavior. Taint tracking usually focuses on the explicit flows of sensitive data, while the information flow analysis may also take into account the implicit flows thus is generally more fine-grained. The frequent use of native code makes the taint tracking and information flow analysis more complicated in both static and dynamic approaches. Dynamic taint tracking \cite{DBLP:conf/osdi/EnckGCCJMS10, DBLP:conf/uss/YanY12, DBLP:conf/dsn/QianLSC14, DBLP:conf/ccs/SunWL16, DBLP:conf/esorics/SchoepeBPS16, DBLP:conf/iwqos/XueQL15} and static information flow analysis \cite{DBLP:conf/ccs/WeiROR14, DBLP:conf/pldi/ArztRFBBKTOM14, DBLP:conf/ndss/GordonKPGNR15, DBLP:conf/iwcmc/LantzJ15, DBLP:conf/eurosp/CalzavaraGM16, DBLP:conf/icse/ArztB16, DBLP:conf/ccs/WeiLOCZ18} have been widely studied to detect privacy leakages of Android apps. These approaches have built different models for the effects of native code, leading to different accuracy.

In the dynamic approaches, TaintDroid \cite{DBLP:conf/osdi/EnckGCCJMS10} tracks the propagation of labeled sensitive data and reports when sensitive data reach the sinks. Its method-level tracking propagates the taints through the JNI call bridges, conservatively specifies that the tainted primitive-type or string arguments of JNI calls can be delivered to taint the returned value. DroidScope \cite{DBLP:conf/uss/YanY12} takes a more fine-grained perspective on inspecting the data flow inside the native instructions but ignores implicit information flow. NDroid \cite{DBLP:conf/dsn/QianLSC14} is built on QEMU as modules that track data flow through JNI. It instruments important JNI functions, e.g., JNI entry/exit, object construction, to track different flows through native contexts, and models the propagation of taint for popular system calls to reduce the performance overhead caused by hooking these frequently invoked functions.

In the static information flow analyses, the state-of-the-art static analyzers \cite{DBLP:conf/ccs/WeiROR14, DBLP:conf/pldi/ArztRFBBKTOM14, DBLP:conf/ndss/GordonKPGNR15, DBLP:conf/eurosp/CalzavaraGM16} do not analyze inside the native component of apps. Instead, they generally resort to some conservative rules to bypass the native calls. These rules can express the taint-propagating relations between used objects, arguments, and return value of native code. Although this kind of abstractions are efficient, they may be neither sound nor precise in modeling taint propagation. Also, the manually crafted models cannot scale up to various third-party native code.

\section{Conclusion and Future Work}\label{sec:conclusion}
In this work, we propose a hybrid framework that combines a control-flow based static binary analysis with a dynamic dependency modeling to build the tainting models of native code in Android apps. Based on such tainting models, we derive fine-grained stubs for the native functions and merge them into the ADI model of information flow analysis engine DroidSafe. With such tainting behavior summaries of native code, the DroidSafe engine can detect sensitive data flows triggered by different types of vulnerabilities of native code. The evaluations have demonstrated the applicability of our approach. Without any intention to depreciate the state-of-the-art inter-language approach, our experimental results emphasize that our approach behaves differently on the accuracy and effectiveness compared with the state-of-the-art inter-language analyzer JN-SAF. Especially, our approach can detect more sensitive flows due to its tight integration with the context-, object- and field-sensitive data-flow analysis. Meanwhile, our approach still confronts performance issues, according to the illustration of Table~\ref{tab:efficiency}.

As future work, we plan to apply our approach to some more efficient analysis framework, e.g., FlowDroid \cite{DBLP:conf/pldi/ArztRFBBKTOM14}. Lightweight symbolic execution and binary-level points-to analysis \cite{DBLP:conf/ndss/KimSZT21} are also expected to improve the efficiency of mutation-based dependency generation or to guide the input choices and reduce false negatives. Moreover, we are endeavoring in parameterizing the disassembly part in our system so that users can substitute IDA with other disassemblers such as \cite{DBLP:conf/codaspy/ZengT18,DBLP:conf/ndss/BaumanLH18,DBLP:conf/uss/Flores-MontoyaS20}.

The source code of \sysname{} has been made publicly available at \url{https://github.com/suncongxd/muDep}.

\ifCLASSOPTIONcaptionsoff
  \newpage
\fi

\bibliographystyle{IEEEtran}

\bibliography{IEEEabrv,mybib}

\vfill

\end{document}